\documentclass[12pt]{iopart}
\usepackage{iopams}
\usepackage{setstack}
\usepackage{graphicx}
\usepackage{cite}
\usepackage[dvipsnames]{xcolor}
\begin{document}

\title[Chemically active 
colloids near osmotic-responsive walls with surface-chemistry 
gradients\hspace*{0.2cm}]
{Chemically active colloids near osmotic-responsive walls with 
surface-chemistry gradients}

\author{M N Popescu, W E Uspal and S Dietrich}

\address{Max-Planck-Institut f\"ur Intelligente Systeme, Heisenbergstr.~3, 
70569 Stuttgart, Germany  \\
IV. Institut f\"ur Theoretische Physik, Universit\"{a}t Stuttgart,
Pfaffenwaldring 57, 70569 Stuttgart, Germany}
\eads{\mailto{popescu@is.mpg.de}, \mailto{uspal@is.mpg.de}, \mailto{dietrich@is.mpg.de}}

\begin{abstract}
Chemically active colloids move by creating gradients in the composition of the 
surrounding solution and by exploiting the differences in their interactions 
with the various molecular species in solution. If such particles move near 
boundaries, e.g., the walls of the container confining the suspension, gradients 
in the composition of the solution are also created along the wall. This 
give rise to chemi-osmosis (via the interactions of the wall with the 
molecular species forming the solution), which drives flows coupling 
back to the colloid and thus influences its motility. Employing an 
approximate ``point-particle'' analysis, we show analytically that -- owing to this 
kind of induced active response (chemi-osmosis) of the wall -- such chemically active 
colloids can align with, and follow, gradients in 
the surface chemistry of the wall. In this sense, these artificial ``swimmers'' 
exhibit a primitive form of thigmotaxis with the meaning of sensing the 
proximity of a (not necessarily discontinuous) physical change in the 
environment. We show that the alignment with the surface-chemistry gradient 
is generic for chemically active colloids as long as they exhibit motility 
in an unbounded fluid, i.e., this phenomenon does not depend on the exact 
details of the propulsion mechanism. The results are discussed in the context of 
simple models of chemical activity, corresponding to Janus particles with 
``source'' chemical reactions on one half of the surface and either ``inert'' or 
``sink'' reactions over the other half. 
\end{abstract}

\vspace{2pc}

%Uncomment for Submitted to journal title message
%\submitto{\JPCM}
%\maketitle
% 

\section{\label{intro}Introduction}

Significant efforts have been made towards the development of small objects 
with sizes in the micrometer range and smaller, which are endowed with 
means of self-propulsion -- and thus to be motile -- in a liquid 
environment \cite{Ismagilov2002,Paxton2004,ozin2005,Paxton2006,solovev2009,
mirkovic2010,Howse2007,Fisher2014,Pine2013,Baraban2012,Wang2012,Sano2010,
Cichos2013,Bechinger2012,Bechinger2014,Stocco2015,Ebbens2010,SenRev2010}. 
These artificial swimmers operate under non-equilibrium conditions such as the  
catalytic promotion of chemical reactions in the surrounding liquid. Their 
motion typically occurs within the regime of very small Reynolds 
($Re$) numbers, and thus they represent potential benchmark examples of 
inertia-less motility mechanisms 
\cite{Purcell1977,Lauga2009,Lauga2010,Spagnolie2012,Yeomans2007,
Yeomans2011,Stark2012prl}. Numerous theoretical 
\cite{Golestanian2005,Ajdari2006,Golestanian2007,Julicher2009,Popescu2010,
Popescu2011,Golestanian2009,Golestanian2012,Seifert2012a,Seifert2012b,
Koplik2013,Crowdy2013,Golestanian2015,Michelin2015,Uspal2015a,Stone2014,
Wurger2013,Wurger2014,MPD16} and numerical 
\cite{Ruckner2007,Tao2008,Kapral2015,deGraaf2015,Lowen2012PNAS,Marchetti2013,
Gompper2015} studies have been devoted to various aspects of the motion of 
self-propelled colloids, ranging from motility mechanisms at the single-particle 
level (``active particles'') 
\cite{Golestanian2005,Ajdari2006,Golestanian2007,Julicher2009,Ruckner2007,Tao2008,
Popescu2010,Popescu2011,Golestanian2009,Golestanian2012,Seifert2012a,Seifert2012b,
Koplik2013,Crowdy2013,Golestanian2015,Michelin2015,Kapral2015,Uspal2015a,
Stone2014,Wurger2013,Wurger2014,deGraaf2015,MPD16} to the emergence of complex 
collective behaviors (``active fluids''), such as `living'' 
crystals \cite{Reinmuler2013,Pine2013,Boquet2012,Bechinger2012}, 
swarming \cite{Prost2007,Marchetti2013,Gompper2015,Boquet2010}, phase separations 
\cite{Prost2007,Masoud2014,Boquet2015,Gompper2015,Cates2015}, or mesoscale 
active turbulence \cite{Lowen2012PNAS,Gompper2015}.

One of the typical experimental realizations of self-propelled particles 
is that of colloids with a surface chemistry designed such as to promote 
catalytically activated chemical reactions in the surrounding liquid environment 
\cite{Howse2007,solovev2009,Wang2012,Ismagilov2002,Paxton2004,ozin2005,Paxton2006,
mirkovic2010,Baraban2012}). These chemically active colloids achieve motility 
through various mechanisms of converting ``chemical'' free energy, obtained from 
locally changing the chemical composition, into mechanical work. Among these 
mechanisms, one often encounters variants, such as diffusio- 
\cite{Howse2007,Baraban2012,Bechinger2012} or electrochemical- 
\cite{Ismagilov2002,Paxton2004,ozin2005,Paxton2006,mirkovic2010} phoresis, 
which exploit the solvent mediated, effective interactions between the surface of 
the active particle and  the reactant and product molecules
\cite{Derjaguin1966,Anderson1989}.  

For chemically active colloids the motion originates from the coupling between 
the distribution of chemical species and the hydrodynamic flow fields produced 
by them \cite{Golestanian2007,Ruckner2007,Julicher2009,Popescu2011}. Both of these 
fields are distorted by nearby interfaces, and therefore it is reasonable to expect 
non-trivial dynamics when active colloids move near confining surfaces. Recent 
reports indeed provide evidence for a very complex behavior. This includes the 
emergence of surface-bound steady-states 
\cite{Uspal2015a,Howse2015,Simmchen2016,Koplik2016} and directed motion due 
to a spatially varying hydrodynamic slip boundary condition \cite{hu15}. If 
the motion occurs near walls, and is in addition exposed to external flows 
or fields, other remarkable features such as rheotaxis of spherical colloids 
\cite{Uspal2015b} and gravitaxis \cite{Ebbens2013,Stark2011,Bechinger2014} appear. 

These phenomena become even richer if the confining surface is by itself 
responsive (in addition to its ``inert'' role of reflecting the chemical and 
hydrodynamic flow fields) to, e.g., the chemical inhomogeneities induced by 
the activity of the colloid. One class of such examples is the case of a 
fluid-fluid interface, the surface tension of which depends on the distribution 
of various molecules across the interfacial region. It has been shown that in such 
cases the chemical activity of the particle can induce Marangoni stresses 
\footnote{Since the surface tension is temperature dependent, a similar 
phenomenon arises if the particle is thermally active, e.g., by being 
heated partially with a laser, rather than chemically active (see, e.g., 
Refs. \cite{Leshansky1997,Wurger2014}).}, 
which for a colloid trapped at the interface lead to self-propulsion along the 
interface \cite{Lauga2012,Stone2014}, or to an effective, long-ranged interaction of 
the colloid with an interface located in its proximity \cite{Alvaro2016}. 

Another class of such surfaces is that of walls on which, in analogy with phoresis, 
chemi-osmotic flows can be induced if the chemical composition of the solution 
varies along the wall \cite{Derjaguin1966,Anderson1989}. This results in 
chemi- (or thermo-) osmotic flows, which extend in the bulk, couple back to the 
colloid, and induce ``drift'' along the surface. Various studies have 
sought to characterize these flows. In order to induce osmotic flows, 
either variations in the composition of the solution have been employed, 
e.g., due to an active colloid in the vicinity of the surface 
(such as in the studies of ``surfers'' in 
Refs. \cite{Reinmuler2013,Pine2013,palacci14,palacci15}), or temperature 
variations along the wall due to hot particles embedded in or glued 
to the surface \cite{Bregulla2016}. 

The motility of self-phoretic active colloids near such surfaces amounts to 
a complex interplay between self-phoresis and the chemi-osmotic response 
of the surface, with the latter strongly depending on the material identity 
of the wall. For example, it has been shown recently that chemical steps 
and stripes of the wall can dock or spatially confine the motion of chemically 
active colloids \cite{Uspal2016}. This naturally leads to the expectation, which 
we shall explore here, that a chemically active colloid may detect and follow -- 
through an induced chemi-osmotic response -- gradual spatial gradients in the 
chemistry of a nearby wall. Such a behavior is a simple form of thigmotaxis, here 
with the meaning of sensing the proximity of a physical change (not necessarily 
discontinuous) in the environment.

In the following we focus on the particular case of a spherical colloid with 
an axisymmetric chemical activity caused by a corresponding catalyst 
distribution. The colloid is considered to be constrained to move with its 
axis parallel to the wall. The wall is responsive via chemi-osmotic flows, and the 
chemistry of the wall varies laterally. We employ a point-particle 
approximation which has been shown \cite{Alvaro2016,Uspal2016} to capture 
the relevant phenomenology in almost quantitative agreement with the 
results obtained from exact numerical calculations. Within this approach 
we obtain simple, physically intuitive expressions for the translation- and 
rotation- contributions for the dynamics of the colloid induced by the 
chemi-osmotic response of the patterned wall. Based on these results we predict 
that in the case of motion near a wall with a constant, linear chemical gradient, 
the active particle will align with the direction of the 
gradient. These general results are then discussed in the context of simple models 
of active colloids defined by suitably selecting the ``activity function'' (see, 
c.f., \Sref{model} for a precise definition). This amounts to prescribing a 
specific chemical reaction at each point on the surface such as to 
mimic either a net production of a solute by the particle, or a particle which 
emits and annihilates the solute so that there is no net production. 

The organization of the paper is as follows. In \Sref{model} we define the 
model, present the governing equations, and derive the contributions of the 
chemi-osmotic response of the patterned wall to the motion of the particle. 
\Sref{res_dis} is devoted to an analysis and a discussion of the emergence 
of thigmotaxis of active particles exhibiting various model activity functions. 
A summary and the conclusions of the study are presented in \Sref{conc}.

\section{\label{model}Model system}

\begin{figure}
 \hspace*{0.1\columnwidth}% 
 \includegraphics[width = 0.2\columnwidth]{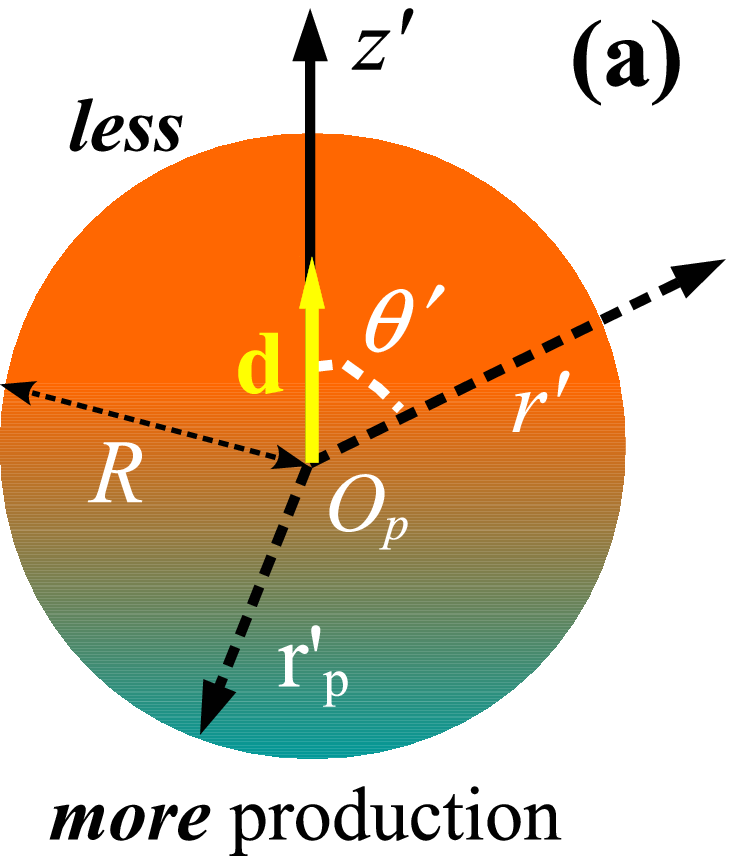}%
 \hspace*{0.1\columnwidth}%
 \includegraphics[width = 0.6\columnwidth]{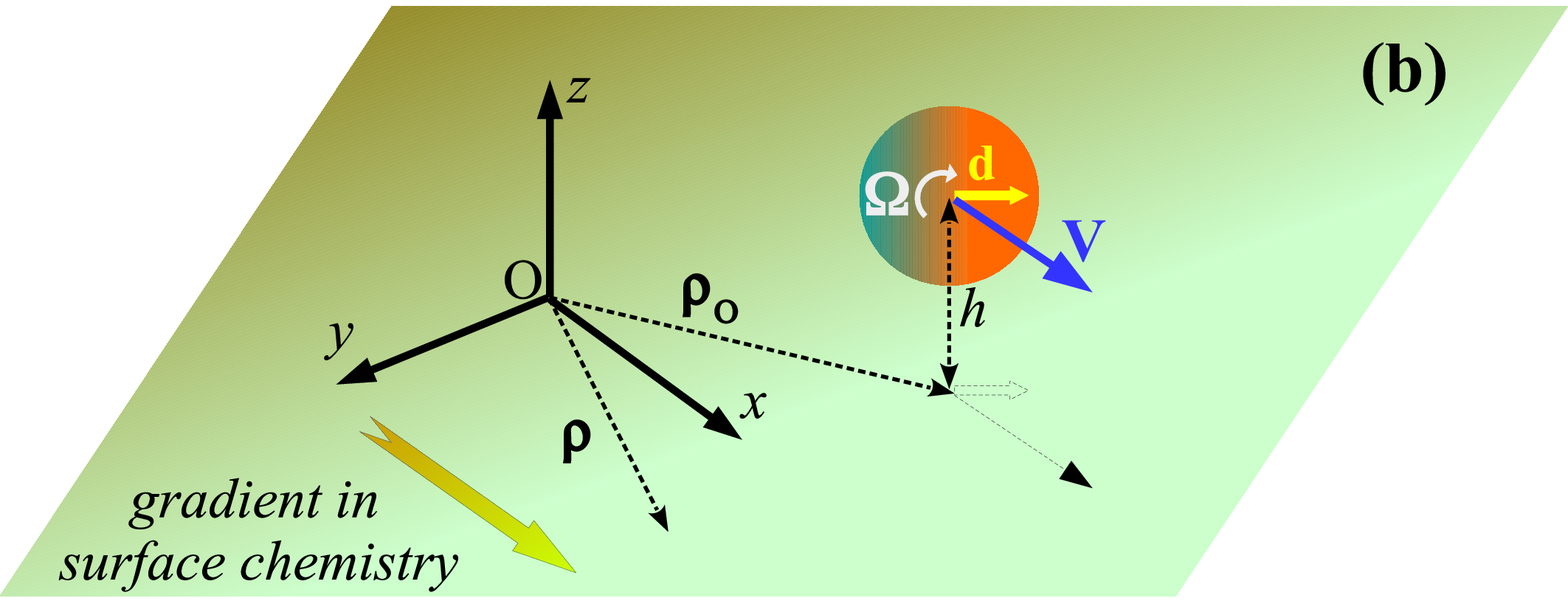}%
\caption{\label{fig1}
(a) Schematic drawing of a spherical, chemically active colloid with an activity 
function which is axisymmetric but lacks fore-aft symmetry. The orientation 
of the colloid is described by the unit vector $\mathbf{d}$ (yellow arrow), which 
points towards the north pole where less solute is produced. $O_p$ denotes 
the center of the particle of radius $R$. (b) An active colloid, as in (a), moving 
laterally with $\mathbf{d} \perp  \mathbf{e}_z$ at a height $h$ (measured from 
the center of the colloid) above the planar wall $(x,y)$ located at $z = 0$. 
The color of the 
plane indicates a linear gradient in the surface chemistry of the wall. Due to 
the chemi-osmotic flow response of the wall, the velocity $\mathbf{V}$ of the 
particle (blue arrow) is not necessarily aligned with $\mathbf{d}$ (which would be 
the case in an unbounded (``bulk'') solution, panel (a)); this is illustrated by 
the thin arrows in the plane $z = 0$ below the particle showing 
the corresponding projection of $\mathbf{d}$ and $\mathbf{V}$, respectively. 
The lateral position of the center of the particle is denoted as 
$\boldsymbol{\rho}_0$.}
\end{figure}

The system of interest is that of a chemically active colloid of radius $R$ 
(\fref{fig1}(a)), moving above a planar wall which has a spatially 
non-uniform surface chemistry, as depicted in \fref{fig1}(b). The colloid is 
suspended in a Newtonian, incompressible solution of viscosity $\mu$. For simplicity, 
we consider the colloid to have a uniform mass density and to be 
density-matched to the solution. The chemical activity of the 
colloid is modeled as the release into (or annihilation from) the solution of 
solute molecules $A$ which are diffusing in the solution with diffusion constant 
$D$. The activity at various parts of the surface of the colloid can be 
different with respect to both the magnitude and/or the source/sink character. 
In \fref{fig1}(a) this is depicted by a variation in 
the color of the surface. Both the colloid and the wall are impermeable to solvent 
and solute. The molecules of species $A$ interact with 
the surface of the colloid, as well as with the surface of the wall. In addition 
to the surfaces being impermeable, these interactions differ from the corresponding 
ones of the solvent molecules and are assumed to be very 
short ranged so that they are significant only within a distance 
$\delta \ll R$ from the surface. The interactions between the $A$ molecules and 
those of the solvent are accounted for by the viscosity $\mu$ of the solution. The 
whole system is in contact with a solute reservoir which prescribes the bulk far 
from the colloid, i.e., the chemical potential and, implicitly, the number 
density $c_\infty$ of species $A$. We further assume that the number density 
${\tilde c}(\mathbf{r})$ of $A$ molecules is everywhere sufficiently small so that 
they can be treated as an ideal gas. 
%%%%%%%%%%%%%%%%%%%%%%%%%%%%%%%%%%%%%%%%%%%%%%%%%%
%%%%%%%%%%%%%%%%%%%%%%%%%%%%%%%%%%%%%%%%%%%%%%%

For molecular species in water-like solvents and for colloids of $\mathrm{\mu m}$ 
in size moving with a velocity of the order of $\mathrm{\mu m/s}$, the 
diffusional relaxation of the solute distribution over distances of the order of $R$ 
is very small. Furthermore, the solution flows with velocities of the same order 
as that of the colloid; thus the Reynolds number $\mathrm{Re} := \rho_m V R/\mu$ 
(with $\rho_m$ the mass density of the solution) and the P{\'e}clet number of the 
solute $\mathrm{Pe}: = R V/D$ are both very small: $\mathrm{Re},~\mathrm{Pe} 
\ll 1$. Therefore the hydrodynamics can be described by the Stokes equations, 
the advective transport of solute can be neglected relative to that by diffusion, 
and the dynamics of ${\tilde c}(\mathbf{r})$ can be analyzed within a quasi-
adiabatic approximation describing the steady-state distribution at the 
instantaneous position of the colloid. 

\subsection{\label{free_space}Chemically active particle in an unbounded fluid}
The motion of such chemically active model colloids in unbounded fluids has been 
studied in detail 
\cite{Golestanian2005,Golestanian2007,Seifert2012a,Koplik2013,Popescu2010}. 
For completeness and convenience here we succinctly summarize the main results. 
As discussed before, we consider only the case of surface activities which 
are axisymmetric; this defines $\mathbf{d}$ (see \fref{fig1}(a)). For some of the 
calculations, it turns out to be convenient to use a system of coordinates (indicated 
by primed quantities) with the origin at the center $O_p$ of the colloidal 
particle and the $z'$-axis oriented along the symmetry axis so that 
$\mathbf{e}_{z'} = \mathbf{d}$. With respect to 
this system, the spherical coordinates $(r',\theta',\phi')$ are introduced in the 
usual way (see \fref{fig1}(a)). 

The excess over the equilibrium, bulk number density of solute 
$c(\mathbf{r'}) := {\tilde c}(\mathbf{r'}) - c_\infty$ obeys the Laplace equation
\begin{equation}
 \label{laplace}
 \nabla'^2 c = 0\,,
\end{equation}
subject to the boundary conditions
\numparts
 \begin{equation}
  \label{c_inf}
  c(|\mathbf{r'}| \to \infty) = 0
 \end{equation}
and
\begin{equation}
\label{def_act}
\left[\mathbf{e}_{r'} \cdot (-D \nabla' c) \right]_{r' = R} = Q f(\theta')\,,
\end{equation}
\endnumparts
where $Q$ (with the units $\mathrm{m^{-2} s^{-1}}$ of a number 
density current) is chosen with the sign convention $Q > 0$ (see below). 
\Eref{def_act} provides the meaning of the ``chemical activity'' for the class 
of models which we consider \footnote{In 
general, it can be expected that the right hand side of equation 
\eref{def_act} takes a more involved form, accounting for the densities of various 
reactant and product species, their different diffusivities, etc.}. It consists 
of a current along the direction $\mathbf{e}_{r'}$, i.e., along the 
direction of the inner normal $\mathbf{n}$ of the surface of the 
particle, which points into the fluid. This current is released into the solution 
at points $\mathbf{r}'_{p} := (R,\theta',\phi')$ on the surface (for 
$\mathbf{r}'_p$ see \fref{fig1}(b)) if $f(\theta') > 0$, and removed from the 
solution if $f(\theta') < 0$. The points with $f(\theta') = 0$ are 
``chemically inert'': at these points equation \eref{def_act} reduces to the 
condition of an impermeable wall. The dimensionless function $f(\mathbf{r}'_p)$, 
which completes the definition of a specific model, is referred to as the 
``activity function''. It encodes all the details related to the kinetics and 
to the mechanisms of the chemical reactions. 

Due to axial symmetry, the solution \cite{Golestanian2007}
\begin{equation}
\label{c_as_series}
 c(r',\theta') = C_0 \sum\limits_{n \geq 0} a_n 
 \left(\frac{R}{r'}\right)^{n+1} P_n(\cos\theta')
\end{equation}
of equations \eref{laplace}, \eref{c_inf}, and \eref{def_act} is independent 
of $\phi'$; $\mathbf{r}'$ denotes a position vector measured from the center $O_p$ 
of the particle, $P_n$ denotes the Legendre polynomial of degree $n$, 
$C_0 := (Q R)/D$ is a characteristic number density, and the dimensionless 
coefficients $a_n$ are given by
\begin{equation}
\label{multipole_mag}
 a_n = \frac{2 n+1}{2 (n+1)} 
 \int\limits_0^\pi \rmd \theta' \sin \theta' f(\theta') P_n(\cos\theta')\,.
\end{equation}
Equation \eref{c_as_series} amounts to a multipole expansion, where the 
first term describes the number density due to a monopole of strength 
$4 \pi Q R^2 a_0$ located at $O_p$, the second term the density due to a dipole of 
strength $4 \pi Q R^3 a_1$ located at $O_p$ and oriented parallel to $\mathbf{d}$, 
etc. We recall that $c(\mathbf{r'})$ is defined relative to the equilibrium 
value $c_\infty$ in the bulk, and thus it can be positive (excess) or 
negative (depletion). In this sense, each of the terms in the expansion 
\eref{c_as_series} of $c(\mathbf{r'})$ (such as the ``dipole'' one 
characterized by the dependence on $\cos \theta'$) carries a specific 
physical meaning. 

Provided that the interaction potential $\Phi(\mathbf{r})$ of the solute 
molecules (relative to that of the solvent molecules 
\cite{Anderson1989,Seifert2012a,Koplik2013}) with the surface of the colloid is 
short ranged as a function of the distance from the surface, this interaction 
is encoded in a so-called phoretic mobility coefficient \cite{Anderson1989}
\begin{equation}
\label{def_b}
 b(\mathbf{r}_p): = (\beta \mu)^{-1} \int\limits_0^\infty \rmd \xi \,\xi \left[ 
\rme^{-\beta \Phi(\mathbf{r}_p + \xi \mathbf{n})} - 1 \right]\,,
\end{equation}
where $\beta$ denotes the inverse of the thermal energy $k_B T$. The dependence 
of $b$ on $\mathbf{r}_p$ exhibits the same symmetries as the 
activity. For the type of active colloids we are focusing on, this 
means that $b(\mathbf{r}_p)$ at a point P on the surface with position 
vector $\mathbf{r}_p$, or $\mathbf{r}'_p$ in the primed system, 
depends only on $\theta'$ (see \fref{fig1}(a)). The effect of the coupling to 
the gradient of the solution composition on the dynamics of the colloid and 
of the solution is accounted for by a so-called phoretic slip hydrodynamic boundary 
condition at the surface of the colloid. 
According to the latter, the hydrodynamic flow $\mathbf{u}$ of the solution is 
prescribed to take a non-zero value at the surface of the colloid 
\cite{Anderson1989,Golestanian2005,Seifert2012a,Koplik2013}:
\begin{equation}
 \label{phor_slip}
  \mathbf{u}(\mathbf{r})|_{\mathbf{r} = \mathbf{r}_p} =  \mathbf{u}_s(\mathbf{r}_p): 
  = - 
 b(\mathbf{r}_p) \nabla_{||} c(\mathbf{r}_p)\,,
\end{equation}
where $\nabla_{||}$ denotes the projection of $\nabla$ onto the tangent plane of the 
surface. The coefficient $b$ can be either positive or negative, depending on 
whether the character of the interaction $\Phi$ is attractive or repulsive; 
for repulsive interactions one has $b < 0$.

The hydrodynamic flow is obtained by solving the incompressible Stokes equations 
subject 
to the boundary condition in \eref{phor_slip} and to the one of a 
quiescent solution at infinity 
(for more details see, e.g., \cite{Blake1971,Michelin2014,HaBr73}). If 
the spherical, chemically active colloid moves in an unbounded 
(\textit{b}ulk) fluid and in the absence of external forces or torques acting 
on it or on the fluid, 
one finds the following simple expression for the velocity 
\cite{Blake1971,Anderson1989,Golestanian2007}:
\begin{equation}
\label{V_bulk}
\mathbf{V}^{(b)} = - (4 \pi R)^{-2} \int\limits_{|\mathbf{r}| = R} \rmd S \,
\mathbf{u}_s(\mathbf{r}_p)\,.
\end{equation} 
In this case the axial symmetry of the system allows only for translation of the 
colloid along the direction defined by $\mathbf{d}$ (see \fref{fig1}(a)). 
If $b$ varies sufficiently weakly across the surface so that $b$ can be taken 
to be a constant, \eref{V_bulk} reduces to the very simple form 
$\mathbf{V}^{(b)} = \frac{2}{3}\, \frac{b Q}{D}\, a_1 \,\mathbf{d}$, which holds 
for any activity function and allows one to identify the velocity scale 
\begin{equation}
 \label{def_V0}
 V_0 : = \frac{|b| Q}{D}\,.
\end{equation}
For Janus colloids, which are half-covered by catalyst, one finds as typical 
experimental values $|\mathbf{V}^{(b)}| \simeq 5~\mathrm{\mu m/s}$ 
\cite{SenRev2010,Ebbens2010,Howse2007,Simmchen2016}. The above simple model with 
uniform $b$, applied to the specific case of ``constant flux'' activity over 
half of the surface while the other half is chemically inert (for which $|a_1| = 3/8$, 
see, c.f., \sref{model_activities}), predicts 
$|\mathbf{V}^{(b)}| = V_0/4$ \cite{Golestanian2007}. This leads to the 
estimate $V_0 \simeq 20~\mathrm{\mu m/s}$. 

\subsection{\label{set-up wall motion}Motion of chemically active particles 
near responsive walls}
We now consider the motion of chemically active colloids near a planar 
wall, located at $z = 0$ (see \fref{fig1} (b)), which is ``responsive'' in a 
sense defined below. The solute number density $c(\mathbf{r})$ obeys the 
Laplace equation \eref{laplace}, subject to the boundary condition far from 
the particle (at infinity) given by \eref{c_inf}, the boundary condition at the 
surface of the particle given by \eref{def_act}, and the additional boundary 
condition of impermeability at the wall
\begin{equation}
 \label{reflect_wall}
 \left[\mathbf{e}_z \cdot (-D \nabla c) \right]_{z = 0} = 0\,.
\end{equation}
(The \textit{interior} normal to the wall is $\mathbf{e}_z$, 
see \fref{fig1} (b).) The solution of the boundary-value problem defined above 
provides the number density $c(\mathbf{r})$ of solute.

As discussed above, besides the impermeability condition the solute molecules 
interact both with the colloid and with the wall. These interactions are 
taken to be short ranged as functions of the distances from the corresponding 
surfaces. Denoting the interaction potential of a molecule $A$ with the wall by 
$\Psi(\mathbf{r})$, the effect of this interaction and of the variations of the 
excess solute number density $c(\boldsymbol{\rho})$ along the wall (where 
$\boldsymbol{\rho}$ is the in-plane position vector, see \fref{fig1} (b)) is, in 
analogy with the situation at the surface of the colloid, a chemi-osmotic shear 
stress along the wall \cite{Derjaguin1966,Anderson1989}. This is accounted for by 
a chemi-osmotic slip hydrodynamic boundary condition at the surface of 
the wall \cite{Derjaguin1966,Anderson1989}:
\begin{equation}
 \label{osmo_slip}
  \mathbf{u}(\mathbf{r})|_{z = 0} = \mathbf{u}_c(\brho): = - 
 b_w(\brho) \nabla_{||} c(\brho)\,,
\end{equation}
with the chemi-osmotic mobility coefficient $b_w$ given by
\begin{equation}
\label{def_osmo_b}
 b_w(\brho): = (\beta \mu)^{-1} \int\limits_0^\infty \rmd \xi \,\xi \left[ 
\rme^{-\beta \Psi(\brho + \xi \mathbf{e}_z)} - 1 \right]\,.
\end{equation}
The slip velocity $\mathbf{u}_c$ (where the subscript ``c'' refers  to 
\textit{c}hemi-osmosis) in \eref{osmo_slip} and \eref{def_osmo_b} defines the 
meaning of a chemi-osmotic response of the wall: if 
the interaction of the solute with the wall reduces to the condition that the wall 
is impermeable, i.e., $\Psi \equiv 0$, one has $b_w \equiv 0$ and $\mathbf{u}_c 
\equiv 0$, which corresponds to the motion near an inert wall. This reduces to 
a no-slip boundary condition at the wall, which solely reflects back the solute 
distribution and the hydrodynamic flow. This was studied in detail in 
\cite{Uspal2015a,Uspal2015b}.

The hydrodynamic flow $\mathbf{u}(\mathbf{r})$ of the solution obeys the 
incompressible Stokes equations
\begin{equation}
\label{Stokes}
 \nabla \cdot \hat{\boldsymbol{\sigma}} = 0\,,\qquad \nabla \cdot \mathbf{u} = 0\,, 
\end{equation}
where $\hat{\boldsymbol{\sigma}}:= -p \hat{\mathbf{\rm I}} + 
\mu \left[\nabla \mathbf{u} 
+ (\nabla \mathbf{u})^{\rm t}\right]$ denotes the stress tensor (of a Newtonian 
fluid), $p$ is the pressure, $\hat{\mathbf{\rm I}}$ is the identity tensor of rank 
three, and the superscript ``t'' denotes the transpose. (We also use the common 
convention that two adjacent vectors (or vector operators) without a dot 
or cross product sign between them denotes the tensorial (dyadic) product.) 
The Stokes equations are to be solved subject to the following boundary 
conditions (at the instantaneous position of the colloid):\newline 
\numparts
- quiescent fluid far from the colloid,
\begin{equation}
\label{quisc_fluid}
 \mathbf{u}(|\mathbf{r}| \to \infty) = 0\,,
\end{equation}
- phoretic slip (known from \eref{phor_slip}) at the surface of the 
colloid (which has rigid body translational and rotational velocities 
$\mathbf{V}$ and $\boldsymbol{\Omega}$, respectively),
\begin{equation}
\label{bc_coll}
\mathbf{u}(\mathbf{r} = \mathbf{r_p}) = \mathbf{V} + \boldsymbol{\Omega} \times 
(\mathbf{r}_p - \mathbf{r}_{O_p}) + \mathbf{u}_s(\mathbf{r}_p) \,,
\end{equation}
where $\mathbf{r}_{O_p}$ denotes the position vector of the center of the 
colloid, and 
\newline 
- osmotic slip (known from \eref{osmo_slip}) at the wall,
\begin{equation}
\label{bc_wall}
\mathbf{u}(\mathbf{r} = \boldsymbol{\rho}) = \mathbf{u}_c(\boldsymbol{\rho}) \,.
\end{equation}
\endnumparts
The boundary value problem posed above is closed by the requirement of 
steady-state motion of the colloid without inertia, i.e., the forces and torques 
acting on the colloid are balanced:
\begin{equation}
 \label{F_T_balance}
 \mathbf{F}_{hyd} + \mathbf{F}_{ext} = 0\,,\qquad 
 \mathbf{T}_{hyd} + \mathbf{T}_{ext} = 0\,,
\end{equation}
where the subscript ``$hyd$'' refers to the hydrodynamic force and torque (due 
to the flow $\mathbf{u}$), while ``$ext$'' refers to any externally imposed forces 
and torques (e.g., due to gravity). 

Depending on the particular system under study here, either $\mathbf{F}_{ext}$ and  
$\mathbf{T}_{ext}$ are given, and the problem is solved for $\mathbf{V}$ and 
$\boldsymbol{\Omega}$, or the velocities $\mathbf{V}$ and 
$\boldsymbol{\Omega}$ are given and the force $\mathbf{F}_{ext}$ and the 
torque $\mathbf{T}_{ext}$ required to achieve that state of motion are calculated. 
(These are often 
called the ``drag'' and the ``force'' problem, respectively \cite{HaBr73}.) Mixed 
problems are also common, in which some of the components of the external 
forces and torques as well as some of the components of the velocities are 
prescribed, all the remaining components being the unknowns for which the problem 
should be solved. 
We consider here, similarly to the discussion in \cite{Uspal2016}, the case 
that the particle motion is constrained to proceed in a plane at height $h$ 
above the wall, with orientation $\mathbf{d} \perp \mathbf{e}_z$ but otherwise 
free to rotate in-plane. While it might be challenging to implement these 
restrictions experimentally (see the discussion in \cite{Uspal2016}), this 
configuration has the advantage of allowing a clear identification of the 
effects due to the 
response of the wall, as well as that of facilitating analytically tractable 
theoretical arguments which lead to closed form, physically intuitive results. 
Thus the hydrodynamics problem to be solved is that of a sphere, with 
phoretic-slip boundary condition at its surface, translating with velocity 
$\mathbf{V} \perp \mathbf{e}_z$ and rotating with angular velocity 
$\boldsymbol{\Omega} = 
\Omega \mathbf{e}_z$ at a fixed height $h$ above a wall with chemi-osmotic slip 
boundary condition. The particle is subject to an external force 
$\mathbf{F}_{ext} = F_{ext} \mathbf{e}_z$, which will be determined by requiring 
that it enforces the constraint $V_z = 0$, and to an external torque 
$\mathbf{T}_{ext} \perp \mathbf{e}_z$, the $x$ and $y$ components of which are 
determined by requiring that they enforce the constraints $\Omega_x = 0$ and 
$\Omega_y = 0$.

Taking advantage of the linearity of the Stokes equations and of the boundary 
conditions, the problem defined above is solved by superposing the 
solutions of the following two sub-problems (see \cite{Uspal2016}). \newline
%%%%%%%%%%%%%%%%%%%%%%%%%%%%%%%%%%%%%%%%%
The \texttt{first sub-problem} (referred to by the label ``sd'') is that of 
\textit{s}elf-\textit{d}iffusiophoresis of a sphere, with phoretic slip 
boundary condition, \eref{bc_coll}, at its surface, and moving near a planar 
wall where the \textit{no-slip} boundary condition holds. In the absence of thermal 
fluctuations 
(which in general are neglected in the present study), owing to the axial 
symmetry of 
the colloid the motion proceeds in the plane which is orthogonal to the wall 
and contains $\mathbf{d}$. We further require that the motion is constrained to 
satisfy $\mathbf{d} \perp \mathbf{e}_z$ and that the center of the particle 
remains located at a height $h$ above the wall. Under 
these conditions, the \textit{s}elf-\textit{d}iffusiophoretic velocity of the 
particle is denoted by $\mathbf{V}^{(sd)}$ which, although not 
explicitly indicated, depends on $h$. These constraints 
are imposed by an external force 
$\mathbf{F}^{(sd)} = F^{(sd)} \mathbf{e}_z$ and an external torque 
$\mathbf{T}^{(sd)} = T^{(sd)}_x \mathbf{e}_x + T^{(sd)}_y 
\mathbf{e}_y$, which are determined by requiring $V^{(sd)}_z = 0$ and 
$\Omega^{(sd)}_x = \Omega^{(sd)}_y = 0$. Due to the symmetry of the system, 
this implies $\mathbf{V}^{(sd)} = V^{(sd)} \mathbf{d}$. This problem was studied 
in detail in 
\cite{Uspal2015a,Simmchen2016,Koplik2016}. Therefore, here we simply note that 
$\mathbf{V}^{(sd)}$ and the constraining quantities $\mathbf{F}^{(sd)}$ and 
$\mathbf{T}^{(sd)}$ can be calculated for a given height $h$ and a 
given phoretic mobility coefficient $b(\mathbf{r}_p)$. In the following, 
we take $V^{(sd)}(h)$ as a parameter which is of the same order of magnitude 
as $V^{(b)}$. \newline
%%%%%%%%%%%%%%%%%%%%%%%%%%%%%%%%%%%%%%%%%%%
The \texttt{second sub-problem} (referred to by the label ``w'' which stands 
for \textit{w}all response) is that of 
a sphere, with no-slip boundary conditions at its surface, translating with 
velocity $\mathbf{V}^{(w)} \perp \mathbf{e}_z$, and rotating with angular 
velocity $\boldsymbol{\Omega}^{(w)} = \Omega_z^{(w)} \mathbf{e}_z$ at a constant 
height $h$ above a wall at which a chemi-osmotic slip boundary condition, 
\eref{bc_wall}, holds. The constraints on the motion are imposed by an external 
force $\mathbf{F}^{(w)} = F^{(w)} \mathbf{e}_z$ and an external torque 
$\mathbf{T}^{(w)} = T^{(w)}_x \mathbf{e}_x + T^{(w)}_y 
\mathbf{e}_y$, which are determined by requiring $V^{(w)}_z = 0$ and 
$\Omega^{(w)}_x = \Omega^{(w)}_y = 0$. The solution of this second sub-problem, 
which provides the contribution of the induced response by the wall to the 
motion of the particle, is the focus of the remainder of the paper. \newline
Before proceeding, we remark that from the solutions of the two sub-problems 
the translational and rotational velocities of the particle follow as 
$\mathbf{V} = \mathbf{V}^{(sd)} + \mathbf{V}^{(w)}$ and 
$\boldsymbol{\Omega} = \boldsymbol{\Omega}^{(w)}$ (because the 
self-diffusiophoresis sub-problem is constrained to not involve rotations), 
respectively. The force $\mathbf{F}_{ext}$ and the torque $\mathbf{T}_{ext}$, 
needed to ensure the desired in-plane motion, are given by the sums of the 
corresponding contributions (``sd'' and ``w'') from the two sub-problems.

\subsection{\label{sec_general_response}The contributions to the motion due to the 
induced response by the wall}

The calculation of the induced response by the wall involves two steps. The first 
one consists of determining the distribution of the solute $c(\boldsymbol{\rho})$ 
at the wall, which is needed in order to find the chemi-osmotic slip 
$\mathbf{u}_c(\boldsymbol{\rho})$, 
\eref{osmo_slip}, for the hydrodynamic boundary condition. The second step 
requires solving, having determined $\mathbf{u}_c(\boldsymbol{\rho})$ 
in the first step, the second hydrodynamics sub-problem formulated in the previous 
subsection. 

Since we seek the qualitative phenomenology and physically intuitive results, 
rather than aiming at a quantitative study, for both steps we employ a 
point-particle (far-field) approximation and keep only few terms in the 
corresponding multipole expansions. (Nevertheless, we keep a sufficient number of 
terms in order to capture the relevant symmetries of the solute distribution at the 
wall, and thus of the wall-driven flows.) Such an approach has been employed 
successfully in previous studies (see, e.g., 
\cite{Spagnolie2012,Alvaro2016} and, in particular, \cite{Uspal2016} which is 
closely related to the present study). In many cases it has turned out that 
this approximation captures very well, 
even semi-quantitatively, the phenomenology for particle center - wall 
distances as small as $h \gtrsim 1.5\,R$ \cite{Uspal2016}.

Within the point-particle approximation, the (excess) density in the presence of the 
wall at $z = 0$ (with reflecting boundary condition, \eref{reflect_wall}, corresponding 
to impermeability) can be constructed straightforwardly from the expansion in 
\eref{c_as_series} by using the method of images: in 
the expansion \eref{c_as_series}, for each source term, located at the position 
$\mathbf{r}_{O_p} = \boldsymbol{\rho}_0 + \mathbf{z}_0$ of the center of the 
particle, an identical copy acting as an image term is placed at 
$\mathbf{r}_{O_p}^* = \boldsymbol{\rho}_0 - \mathbf{z}_0$. This ensures that the 
boundary condition \eref{reflect_wall} is satisfied by the sum $c(\mathbf{r'}) + 
c_{image}(\mathbf{r'})$ of the source and image contributions. Here 
$\boldsymbol{\rho}_0 = x_0 \mathbf{e}_x + y_0 \mathbf{e}_y$ 
is the position vector of the projection of $\mathbf{r}_{O_p}$ onto the plane 
$z = 0$, $\mathbf{z} = z \mathbf{e}_z$, and $\mathbf{z}_0 = h \mathbf{e}_z$ (see 
\fref{fig1}). With $c(\mathbf{r'}) = c(\mathbf{r}-\mathbf{r}_{O_p}) = c(\boldsymbol{\rho} 
- \boldsymbol{\rho}_0 +\mathbf{z} - \mathbf{z}_0)$ and $c_{image}(\mathbf{r'}) = 
c(\boldsymbol{\rho} - \boldsymbol{\rho}_0 +\mathbf{z} + \mathbf{z}_0)$, due to 
$|\boldsymbol{\rho} - \boldsymbol{\rho}_0 - \mathbf{z}_0| = |\boldsymbol{\rho} - 
\boldsymbol{\rho}_0 + \mathbf{z}_0|$, it follows that the solution 
$c(\boldsymbol{\rho},z=0)$ fulfilling the boundary condition at 
the point $\boldsymbol{\rho} = x \mathbf{e}_x + y \mathbf{e}_y$, i.e., $z = 0$, 
is twice the expression in \eref{c_as_series} evaluated at the wall, 
i.e., at $\mathbf{r'} = \boldsymbol{\rho} - \boldsymbol{\rho}_0 - \mathbf{z}_0$:
\begin{equation}
\label{wall_dens}
c(\boldsymbol{\rho}) = C_0 \sum_{n \ge 0} c^{(n)} (\boldsymbol{\rho})\,,
\end{equation}
with the \newline
\numparts
- \textit{m}onopole ($n = 0$) contribution
\begin{equation}
\label{dens_mon}
c^{(m)} (\boldsymbol{\rho}) := 2 a_0 
\frac{R}{|\boldsymbol{\rho} - \boldsymbol{\rho}_0 - \mathbf{z}_0|}\,,
\end{equation}
- \textit{d}ipole ($n = 1$) contribution
\begin{equation}
\label{dens_dip}
c^{(d)} (\boldsymbol{\rho}) := 2 a_1 
\left(\frac{R}{|\boldsymbol{\rho} - \boldsymbol{\rho}_0 - \mathbf{z}_0|}\right)^2 
\cos \theta' \,,
\end{equation}
- \textit{q}uadrupole ($n = 2$) contribution
\begin{equation}
\label{dens_quad}
c^{(q)} (\boldsymbol{\rho}) :=  a_2 
\left(\frac{R}{|\boldsymbol{\rho} - \boldsymbol{\rho}_0 - \mathbf{z}_0|}\right)^3  
\left(3 \cos^2 \theta' - 1 \right)\,,
\end{equation}
 - \textit{o}ctopole ($n = 3$) contribution
\begin{equation}
\label{dens_oct}
c^{(o)} (\boldsymbol{\rho}) :=  a_3 
\left(\frac{R}{|\boldsymbol{\rho} - \boldsymbol{\rho}_0 - \mathbf{z}_0|}\right)^4  
\left(5 \cos^2 \theta' - 3 \right) \cos \theta'\,,~ etc.
\end{equation}
\endnumparts
The angle $\theta'$ between $\mathbf{d}$ and $\mathbf{r'}$ is given by
\begin{equation}
 \label{cos_th}
 \cos \theta' = \frac{\mathbf{d} \cdot \mathbf{r'}}{|\mathbf{r'}|} 
\overset{(\mathbf{d}\perp \mathbf{e}_z)}{=} 
\mathbf{d} \cdot \frac{\boldsymbol{\rho} - 
 \boldsymbol{\rho}_0}{|\boldsymbol{\rho} - \boldsymbol{\rho}_0 -\mathbf{z}_0|}\,.
\end{equation}

%%%%%%%%%%%%%%%%%%%%%%%%%%%%%%%%%%%%%%%%%%%%%%%%%%
\begin{figure}[!htb]
\flushright{
\includegraphics[width = 0.8\columnwidth]{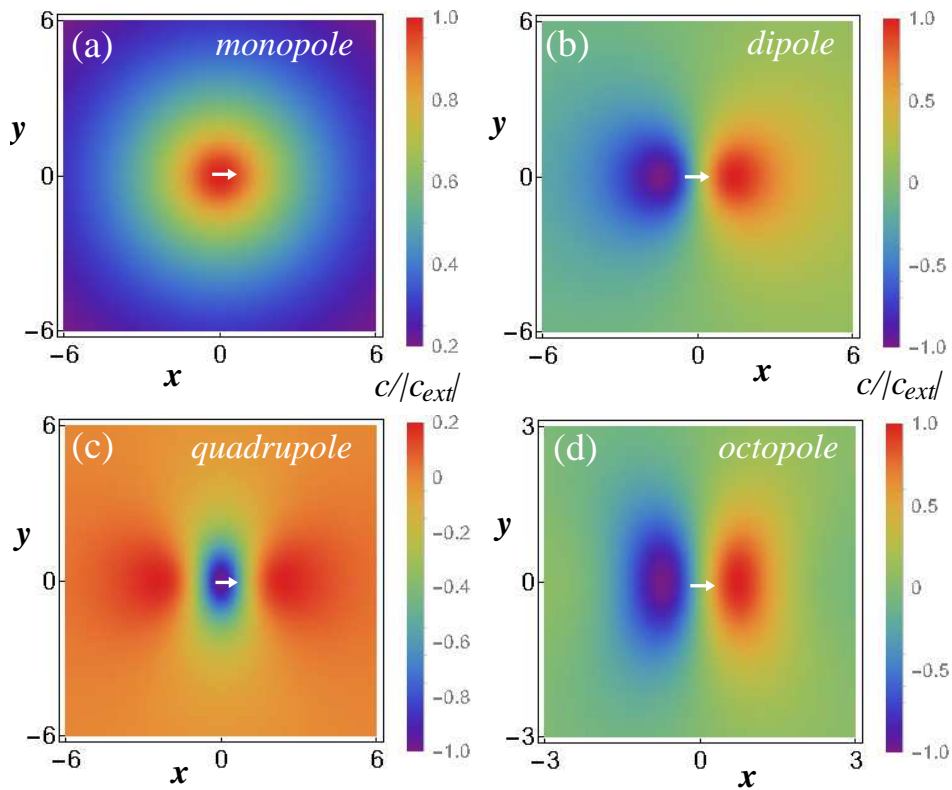}
}%
\caption{\label{fig2}
Various density distributions $c(\boldsymbol{\rho}-\boldsymbol{\rho}_0)$ 
at the wall $z = 0$ (color coded) in units of the absolute values 
$|c_{ext}|$ of their corresponding extrema. The panels show the monopole, 
dipole, quadrupole, and octopole contributions (see \eref{dens_mon} - 
\eref{dens_oct}) of an active particle located at $h = 2 R$ above the 
lateral position $\boldsymbol{\rho}_0$. All distances are measured in units 
of the particle radius $R$. The white arrows indicate the projections 
$\mathbf{d}_{||}$ of $\mathbf{d}$ onto the plane $z = 0$. The considered 
ranges of values (see the color bar to the right of each plot) differ 
between the four panels. For reasons of clarity panel (d) shows only a 
smaller region around the lateral position $\boldsymbol{\rho}_0$ of the 
center $O_p$ of the particle.
}
\end{figure}
%%%%%%%%%%%%%%%%%%%%%%%%%%%%%%%%%%%%%%%%%%%%%%%
In \fref{fig2} we illustrate each multipole contribution, \eref{dens_mon} 
- \eref{dens_oct}, to the excess number density at the wall for the particular value 
$h = 2\,R$. It can be seen that, as borne out by \eref{dens_mon}, the monopole 
contribution is the \textit{only} one with in-plane radial symmetry around the projection 
of the particle center $P_O$ onto the plane $z = 0$. All the higher order ones 
exhibit in-plane 
mirror symmetry with respect to the line $\mathbf{d}_{||}$ (white arrows in 
\fref{fig2}) which is parallel to $\mathbf{d}$. Intuitively, one thus infers that 
if the 
chemical properties of the wall are anisotropic and exhibit a specific
direction $\mathbf{d'}$ (as in, e.g., \fref{fig1} (b)), which happens to 
be distinct from $\mathbf{d}$, the various terms in the expansion may cause 
qualitatively different effects. The monopole contribution perceives 
``passively'' the direction $\mathbf{d'}$ because this distribution, and therefore 
the response induced at the wall, is independent of the specific 
orientation of $\mathbf{d}_{||}$, i.e., of how the 
particle is oriented with respect to the anisotropy in the properties of the wall. 
The magnitude of $\nabla_{||}c^{(m)}$ depends only on the distance 
$|\boldsymbol{\rho} - \boldsymbol{\rho}_0|$ from the lateral position 
$\boldsymbol{\rho}_0$. However, since $b_w$ varies along $\mathbf{d'}$ the 
magnitude of the chemi-osmotic slip depends on the position $\boldsymbol{\rho} - 
\boldsymbol{\rho}_0$, and not only on this distance. Therefore, along the 
direction $\mathbf{d'}$ the response 
of the wall, in the form of chemi-osmotic flows, does not have fore-aft 
symmetry. Hence it induces a translation of the colloid along $\mathbf{d'}$. 
This translation is independent of the orientation of $\mathbf{d}_{||}$ relative 
to that of $\mathbf{d'}$. However, the higher order terms offer the means for 
competing effects and the 
interplay between the ``wall preferred'' symmetry characterized by $\mathbf{d'}$ and 
the ``density preferred''  one $\mathbf{d}_{||}$. Since the particle is allowed 
to undergo in-plane rotations, $\mathbf{d}$ is actually free to rotate. This can 
lead to the selection of certain orientations of $\mathbf{d}$ with respect to 
$\mathbf{d'}$ so that an ``active'' sensing of $\mathbf{d'}$ emerges. This point will 
be analyzed in detail in, c.f., \sref{res_dis}. We also remark that with respect to 
the line $\mathbf{d}_\perp$ in the plane of the wall, which passes through 
the projection $P_O$ and is perpendicular to $\mathbf{d}_{||}$, the even-$n$ 
multipole contributions are mirror-symmetric (fore-aft symmetry), while the odd-$n$ ones 
are mirror-antisymmetric (i.e., corresponding to reflection and a change in 
sign), and thus their fore-aft symmetry is broken. The latter implies that 
there is, e.g., 
a response which has also a directionality with respect to $\mathbf{d}_{||}$ 
(i.e., it can give rise to \textit{surfing} \cite{palacci15}). Finally, as can be 
easily inferred from \eref{dens_mon} - \eref{dens_oct}, the magnitude of the 
contributions decreases steeply with increasing order $n$, and thus the 
first few terms dominate the behavior of the density $c(\boldsymbol{\rho})$, 
\eref{wall_dens}, for in-plane distances $|\boldsymbol{\rho} - 
\boldsymbol{\rho}_0|/R \gtrsim 3$.

We now proceed to the solution of the second hydrodynamic sub-problem. 
Since here only the translational velocity $\mathbf{V}^{(w)}$ and the angular 
velocity $\boldsymbol{\Omega}^{(w)}$ of the particle are of particular 
interest, we approach this hydrodynamic sub-problem by using the 
Lorentz reciprocal theorem \cite{Lorentz_original,Lorentz_transl,HaBr73,KiKa91}. 
This states that, in the absence of volume forces, any two solutions 
$\mathbf{u}_0(\mathbf{r})$ and $\mathbf{u}_a(\mathbf{r})$ of the incompressible 
Stokes equations within the same domain $\cal D$, i.e., solutions subject to 
different boundary conditions but on the very same boundaries $\partial 
\cal D$, obey the relation
\begin{equation}
\int\limits_{\partial \cal D}^{} \mathbf{u}_0 \cdot 
\hat{\boldsymbol{\sigma}}_a \cdot \mathbf{n} \,\rmd S 
= \int\limits_{\partial \cal D}^{} \mathbf{u}_a \cdot 
\hat{\boldsymbol{\sigma}}_0 \cdot \mathbf{n} \,\rmd S\,,
\label{rec-theo-1}
\end{equation}
where $\hat{\boldsymbol{\sigma}}_0$ and $\hat{\boldsymbol{\sigma}}_a$ 
denote the stress 
tensors corresponding to the two flow fields. We choose for the problem ``0'' the above 
second sub-problem. For the so-called \textit{a}uxiliary problem ``a'', 
we consider a sphere with radius $R$ and no slip at its surface which 
is driven by certain forces $\mathbf{F}_a$ and torques $\mathbf{T}_a$. It moves 
with constant translational velocity $\mathbf{V}_a$ and angular 
velocity $\boldsymbol{\Omega}_a$ through a fluid at the same height $h$ from a 
planar wall with \textit{no slip} boundary condition:
\begin{equation}
 \label{no_slip_wall}
\mathbf{u}_a|_{z = 0} = 0\,. 
\end{equation}
The fluid is assumed to be quiescent far away from the particle, i.e., 
$\mathbf{u}_a(|\mathbf{r}| \to \infty) = 0$.

For our system the boundary $\partial \cal D$ consists of the surface of the 
sphere, the wall, and the surface $S_\infty$ at infinity, e.g., a hemisphere 
(in the half plane $z > 0$) with a radius $R_\infty \to \infty$. Since 
both $\mathbf{u}_0(\mathbf{r})$ and $\mathbf{u}_a(\mathbf{r})$ decay at least 
$\propto\, |\mathbf{r}|^{-1}$ for $|\mathbf{r}| \to \infty$, the contributions 
from the integrals over $S_\infty$ are vanishingly small. Thus only the 
integrals over the surface of the sphere and over the wall contribute to 
\eref{rec-theo-1}. By replacing the flows at the surfaces by the expressions 
for the corresponding boundary conditions (in the auxiliary problem, the 
boundary condition at the sphere has the same form as \eref{bc_coll} but with 
$\mathbf{V}_a$ and $\boldsymbol{\Omega}_a$ replacing the corresponding quantities 
and with $\mathbf{u}_s = 0$), by using that 
$\mathbf{V}^{(w)}$, $\boldsymbol{\Omega}^{(w)}$, $\mathbf{V}_a$, and 
$\boldsymbol{\Omega}_a$ are constant vectors (which allows one to factor them 
out of the integrals), and by noting that in each problem the hydrodynamic forces 
and torques must be exactly balanced by the forces and torques driving the colloid,
one arrives at (for $\mathbf{u}_c$ see \eref{osmo_slip})
\begin{equation}
 \label{system_from_reciproc}
\mathbf{V}^{(w)} \cdot \mathbf{F}_a + 
\boldsymbol{\Omega}^{(w)} \cdot \mathbf{T}_a - \mathbf{V}_a \cdot 
\mathbf{F}^{(w)} - \boldsymbol{\Omega}_a \cdot \mathbf{T}^{(w)} 
= \int\limits_{z = 0} \rmd S \,\mathbf{u}_c(\boldsymbol{\rho}) \cdot 
\hat{\boldsymbol{\sigma}}_a \cdot \mathbf{e}_z\,.\qquad
\end{equation}
Since, by construction of the second sub-problem, $\mathbf{F}^{(w)}$ has only a 
$z$-component, while $\mathbf{T}^{(w)}$ has no $z$-component, for auxiliary 
problems which involve only translations along the $x$- or $y$-axis and/or rotations 
around the $z$-axis it follows that the last two terms on the left hand 
side of \eref{system_from_reciproc} vanish. In such cases, we obtain 
equations involving only the desired unknowns, i.e., $\mathbf{V}^{(w)}$ and 
$\boldsymbol{\Omega}^{(w)}$. Accordingly, we select the following three auxiliary 
problems, indexed by $a = 1,2,3$:\newline
\indent \textbullet~$a = 1$: sphere translating along the $x$-axis (at $z = h$) due 
to 
a force $\mathbf{F} = F \mathbf{e}_x$;\newline
\indent \textbullet~$a = 2$: sphere translating along the $y$-axis (at $z = h$) 
due to 
a force $\mathbf{F} = F \mathbf{e}_y$;\newline
\indent \textbullet~$a = 3$: sphere spinning around the $z$-axis (at $z = h$) 
due to a torque $\boldsymbol{\tau} = \tau \mathbf{e}_z$.\newline
In the first two auxiliary problems, due to the viscous friction on the sphere 
being larger on the side closer to wall, a rotation (rolling) occurs. 
Thus also a torque must be applied to the particle in order to achieve a 
purely translational motion. However, it turns out that this translation-rotation 
coupling is negligibly small unless the motion takes place very close to the 
wall, i.e., for $h/R \lesssim 1.1$ \cite{Goldman1967} (see also 
\cite{Uspal2015a,Uspal2016}), which is outside the range we are interested in. 
Therefore, in the first two auxiliary problems one can neglect these torques, 
which leads to a decoupling of the three equations (i.e., 
\eref{system_from_reciproc} with $a = 1,2,3$) and to the simple result
\numparts
\begin{equation}
 \label{Vx_form1}
 V^{(w)}_x = \int\limits_{z = 0} \rmd S \,\mathbf{u}_c(\boldsymbol{\rho}) \cdot 
\frac{\hat{\boldsymbol{\sigma}}_1}{F} \cdot \mathbf{e}_z\,,
\end{equation}
\begin{equation}
 \label{Vy_form1}
 V^{(w)}_y = \int\limits_{z = 0} \rmd S \,\mathbf{u}_c(\boldsymbol{\rho}) \cdot 
\frac{\hat{\boldsymbol{\sigma}}_2}{F} \cdot \mathbf{e}_z\,,
\end{equation}
 \begin{equation}
 \label{Oz_form1}
 \Omega^{(w)} = \int\limits_{z = 0} \rmd S \,\mathbf{u}_c(\boldsymbol{\rho}) \cdot 
\frac{\hat{\boldsymbol{\sigma}}_3}{\tau} \cdot \mathbf{e}_z\,
\end{equation}
\endnumparts
with $\mathbf{u}_c$ given by \eref{osmo_slip}.

For the stress tensors $\hat{\boldsymbol{\sigma}}_a$ corresponding to the 
auxiliary problems we use a point particle approximation. This is similar to the 
approach we have taken for calculating the density of solute at the wall. Under this 
approximation, the auxiliary problems reduce to those of the flows induced  by
point forces or torques located at $\mathbf{r}_{O_p}$, which have been 
calculated previously \cite{Blake1971}. Due to the structure of \eref{Vx_form1} - 
\eref{Oz_form1}, according to which  
$\mathbf{u}_c(\boldsymbol{\rho})$ has only the in-plane $x$- and $y$- 
components, 
only the components ${\hat{\boldsymbol{\sigma}}}_{a,xz}$ and 
${\hat{\boldsymbol{\sigma}}}_{a,yz}$ are needed. At the wall, the 
so-called ``tractions'' $\mathbf{g} := 
\hat{\boldsymbol{\sigma}} \cdot \mathbf{e}_z$ are given by \cite{Uspal2016}\newline
\numparts 
\indent \textbullet~$a = 1$ (point force $\mathbf{F} = F \mathbf{e}_x$):
\begin{equation}
 \label{point_force_prob1}
\mathbf{g}_1:= \frac{\hat{\boldsymbol{\sigma}}_1}{F} \cdot \mathbf{e}_z 
= -\frac{3 h}{2 \pi} (x-x_0) \frac{\boldsymbol{\rho} - \boldsymbol{\rho}_0}
 {|\boldsymbol{\rho} - \boldsymbol{\rho}_0 -\mathbf{z}_0|^5}\,,\quad
\end{equation}

\indent \textbullet~$a = 2$ (point force $\mathbf{F} = F \mathbf{e}_y$):
\begin{equation}
 \label{point_force_prob2}
\mathbf{g}_2:= \frac{\hat{\boldsymbol{\sigma}}_2}{F} \cdot \mathbf{e}_z =
 -\frac{3 h}{2 \pi} (y-y_0) \frac{\boldsymbol{\rho} - \boldsymbol{\rho}_0}
 {|\boldsymbol{\rho} - \boldsymbol{\rho}_0 -\mathbf{z}_0|^5}\,,\quad 
\end{equation}

\indent \textbullet~$a = 3$ (point torque $\boldsymbol{\tau} = \tau \mathbf{e}_z$):
\begin{equation}
 \label{point_torque_prob3}
\mathbf{g}_3:= \frac{\hat{\boldsymbol{\sigma}}_3}{\tau} \cdot \mathbf{e}_z =
 -\frac{3 h}{4 \pi} \frac{(y-y_0) \mathbf{e}_x - (x-x_0) \mathbf{e}_y}
 {|\boldsymbol{\rho} - \boldsymbol{\rho}_0 -\mathbf{z}_0|^5}\,.\quad  
\end{equation}
\endnumparts
The set of equations \eref{system_from_reciproc}, \eref{Vx_form1} - \eref{Oz_form1}, 
\eref{wall_dens}, \eref{dens_mon} - \eref{dens_quad}, and \eref{osmo_slip} completes 
the calculation of the response induced by the wall. It 
provides $\mathbf{V}^{(w)}$ and $\boldsymbol{\Omega}^{(w)}$ in terms of 
a series of contributions arising from the corresponding multipole  
terms in the expansion of the density at the wall. In the following we restrict the 
discussion to the calculation of the angular velocity because it determines 
the emergence of thigmotaxis, the phenomenon which is the focus of the 
present study. 

\section{\label{res_dis}Results and discussion} 

The dependence of the chemi-osmotic mobility coefficient $b_w(\boldsymbol{\rho})$, 
\eref{def_osmo_b}, at the lateral position $\boldsymbol{\rho}$ on the wall 
reflects the chemical patterns occurring at the wall. The coefficient 
$b_w(\boldsymbol{\rho})$ must be a bounded function. However, in the following 
we shall consider also  a linear dependence on $|\boldsymbol{\rho}|$; this should 
be interpreted  physically as the case of a plate with a lateral extent being much 
larger than $R$ or $h$, yet finite. 
This case is appealing due to its conceptual simplicity (similar to studies of 
rheotaxis 
which are employing a planar shear flow). It captures, as shown below, the alignment 
of the particle axis with the direction of the surface gradient. This allows one 
to follow the emergence of thigmotaxis without carrying out cumbersome algebra.

Before proceeding with specific choices for the spatial variations of 
the chemistry of the 
wall and for the activity function $f(\mathbf{r}'_P)$, we note that the steep decay 
of the tractions $\mathbf{g}_k,~k = 1,2,3$, as functions of 
$|\boldsymbol{\rho}-\boldsymbol{\rho}_0|$ allows one to 
re-write the integrals in \eref{Vx_form1} - \eref{Oz_form1} in a more convenient 
form:
\begin{eqnarray}
 I_k &:= -\int\limits_{z = 0} \rmd S \, b_w(\boldsymbol{\rho}) 
 \nabla_{||} c(\boldsymbol{\rho}) \cdot \mathbf{g}_k \nonumber \\
 & = 
 -\int\limits_{z = 0} \rmd S \,  
 \nabla_{||} \cdot \left[b_w(\boldsymbol{\rho}) c(\boldsymbol{\rho}) 
 \mathbf{g}_k \right] + \int\limits_{z = 0} \rmd S \,  
 c(\boldsymbol{\rho}) \nabla_{||} \cdot \left[b_w(\boldsymbol{\rho}) 
 \mathbf{g}_k \right]\nonumber\\
 &= -\lim_{\rho_\infty \to\,\infty} 
 \oint\limits_{{\cal C}_\infty} \,\rmd \phi \, \rho_\infty  
 b_w(\boldsymbol{\rho}) c(\boldsymbol{\rho}) (\mathbf{g}_k \cdot \mathbf{e}_\rho) 
 \nonumber + \int\limits_{z = 0} \rmd S \,  
 c(\boldsymbol{\rho}) \nabla_{||} \cdot \left[b_w(\boldsymbol{\rho}) 
 \mathbf{g}_k \right]\,,\nonumber
 \end{eqnarray}
where $\rho_\infty$ denotes the radius of the contour ${\cal C}_\infty$ in the 
plane $z = 0$. Since $g_k(\rho \to \infty) \sim \rho^{-3}$ and $c(\rho \to \infty) 
\sim \rho^{-1}$, 
while at most $b_w(\rho \to \infty) \sim \rho$, the contour integral 
diminishes at least $\propto$ $\rho_\infty^{-2}$ and therefore vanishes in the limit 
$\rho_\infty \to\,\infty$. 
Therefore
\begin{equation}
\label{I_k}
 I_k =  \int\limits_{z = 0} \rmd S \,  
 c(\boldsymbol{\rho}) \left[(\nabla_{||} b_w(\boldsymbol{\rho}))\cdot \mathbf{g}_k 
 + b_w(\boldsymbol{\rho}) (\nabla_{||} \cdot \mathbf{g}_k) \right]\,.
\end{equation}
This form provides particular insight into the calculation of the 
angular velocity $\Omega = I_3$, \eref{Oz_form1} and \eref{I_k}, which determines 
the rotation of the particle axis $\mathbf{d}$. (We recall that, 
as discussed at the end of \sref{set-up wall motion}, self-diffusiophoresis does 
not 
contribute to the in-plane rotations of the particle, i.e., $\Omega = 
\Omega^{(w)}$.) 
Since $\nabla_{||} \cdot \mathbf{g}_3 \equiv 0$ (which follows from 
\eref{point_torque_prob3} by direct calculation), one arrives at 
\begin{equation}
\label{Oz_form2}
 \Omega =  \int\limits_{z = 0} \rmd S \,  
 c(\boldsymbol{\rho}) \, \mathbf{g}_3 \cdot \nabla_{||} b_w(\boldsymbol{\rho})  \,.
\end{equation}
From \eref{Oz_form2} it immediately follows that in the case of a spatially 
homogeneous 
chemistry of the wall (i.e., $b_w(\boldsymbol{\rho}) = const$) the angular velocity $\Omega$ 
vanishes identically (as it should, see the corresponding discussion in \sref{set-up 
wall motion}). 
Thus in this case the induced response of the wall, i.e., the chemi-osmotic 
driven flow, cannot give rise to rotations of the axis of the active colloid. 

\subsection{Rotation in response to a constant surface-chemistry gradient}
We now turn to the focus of the present study, which is a spatially constant 
gradient of the surface chemistry, as indicated in \fref{fig1}(b). We 
choose the coordinate system such that the $x$-axis coincides with the direction 
of the gradient. The orientation $\mathbf{d}_{||}$ of the particle can be expressed 
in terms of the corresponding orthogonal unit vectors: 
\begin{equation}
 \mathbf{d}_{||} = \cos \psi \,\mathbf{e}_x + \sin \psi \,\mathbf{e}_y\,.
\end{equation}
Within this model, the chemi-osmotic mobility coefficient 
$b_w(\boldsymbol{\rho})$ 
takes the form
\begin{equation}
 \label{linear_grad}
b_w(\boldsymbol{\rho}) \equiv b_w(x,y) = b_0 + \Delta \, x\,, \qquad \Delta > 0\,,
\end{equation}
which implies $b_w(0) = b_0$. In this case one has $\nabla_{||} b_w(\boldsymbol{\rho}) 
= \Delta \, \mathbf{e}_x$. From \eref{Oz_form2} and \eref{point_torque_prob3} it 
follows that 
\begin{equation}
\label{Oz_lin_grad}
 \Omega =  - \frac{3 h \Delta}{4 \pi} \int\limits_{-\infty}^\infty \, \rmd x \, 
 \int\limits_{-\infty}^\infty \,  \rmd y \,
 c(\boldsymbol{\rho}) \, \frac{(y-y_0)}{|\boldsymbol{\rho} - \boldsymbol{\rho}_0 - 
 \mathbf{z}_0|^5}\,  
 \,.
\end{equation}
The ratio in the last term of the integrand is an odd function of $y-y_0$, and each term 
in the expansion \eref{wall_dens} of $c(\boldsymbol{\rho})$ has parity with respect to 
$y-y_0$ (see, e.g., \eref{dens_dip} - \eref{dens_quad}). Therefore, only the terms in 
the expansion \eref{wall_dens} which are odd functions of $y-y_0$ contribute to 
$\Omega$; 
these terms correspond to odd values of the index $n$ in \eref{wall_dens}. 

Considering the first four terms (\eref{dens_dip} - \eref{dens_quad}) in the expansion 
of the density at the wall (see \sref{sec_general_response}), the above arguments 
imply that the monopole and quadrupole terms do not give rise to a rotation of the 
axis of the colloid. The angular velocity contribution due to the \textit{d}ipole 
term is given by
\begin{eqnarray}
 &\Omega^{(d)}  :=  - \frac{3 h \Delta }{4 \pi} \int\limits_{z=0}\, \rmd S \, 
  c^{(d)}(\boldsymbol{\rho}) \, \frac{(y-y_0)}{|\boldsymbol{\rho} - 
  \boldsymbol{\rho}_0 - \mathbf{z}_0|^5}\qquad\nonumber\\
 & 
 = -
 %\quad\overset{\eref{dens_dip},\,\eref{cos_th}}{=} - 
 \frac{3 a_1 C_0 R^2 h \Delta}{2 \pi} 
 \int\limits_{z=0}\, \rmd S \, 
 \frac{(x-x_0)(y-y_0) \cos\psi + (y-y_0)^2 \sin\psi}{|\boldsymbol{\rho} - 
 \boldsymbol{\rho}_0 - \mathbf{z}_0|^8}\nonumber\\
 &\quad\quad = - \frac{3 a_1 C_0 R^2 h \Delta \sin\psi}{4 \pi} \int\limits_{z=0}\, 
 \rmd S \, 
 \frac{|\boldsymbol{\rho} - \boldsymbol{\rho}_0|^2}{|\boldsymbol{\rho} - 
 \boldsymbol{\rho}_0 - \mathbf{z}_0|^8}\,,\nonumber
 \end{eqnarray}
 where in passing from the first line to the second we have used \eref{dens_dip} and 
\eref{cos_th}, and in passing  
 from the second to the 
third line we have used the fact that the 
first product 
in 
 the nominator has odd parity with respect to both $(x-x_0)$ and $(y-y_0)$ and thus 
gives a 
 vanishing contribution upon integration, while the integral of the second term is 
 invariant with respect to the replacement $(y-y_0) \to (x-x_0)$. This leads to
\begin{equation}
\label{Oz_dip}
 \frac{\Omega^{(d)}}{\Omega_0} = 
 -\frac{1}{8} a_1 \left(\frac{R}{h}\right)^3 \sin \psi\,,
\end{equation}
with the characteristic angular velocity
\begin{equation}
 \label{char_Oz}
\Omega_0 := \frac{C_0 \Delta}{R} = \frac{Q \Delta}{D}  > 0\,.
\end{equation}
The contribution to the angular velocity due to the \textit{o}ctopole term 
is given by
\begin{equation}
 \Omega^{(o)}  :=  - \frac{3 h \Delta}{4 \pi} \int\limits_{z=0}\, \rmd S \, 
 c^{(o)}(\boldsymbol{\rho}) \, \frac{(y-y_0)}{|\boldsymbol{\rho} - \boldsymbol{\rho}_0 - 
 \mathbf{z}_0|^5} \,.
\end{equation}
The calculation of the integral is lengthy, but it involves steps similar to those 
leading to \eref{Oz_dip}. Therefore we provide only the final result:
\begin{equation}
\label{Oz_oct}
 \frac{\Omega^{(o)}}{\Omega_0} = \frac{3}{64} a_3 \left(\frac{R}{h}\right)^5 \sin \psi\,.
\end{equation}
Adding these contributions, one arrives at the following expression for the angular 
velocity due 
to the induced response at the wall:
\begin{equation}
\label{Oz_total}
 \frac{\Omega}{\Omega_0} = \frac{1}{8} \left(\frac{R}{h}\right)^3 
 \left[-a_1 + \frac{3}{8} a_3 \left(\frac{R}{h}\right)^2 + 
 {\cal O}\left( \left(\frac{R}{h}\right)^4 \right) \right] \sin \psi\,.
\end{equation}

The in-plane rotation of the axis $\mathbf{d}$ of the colloid due to the 
above contributions thus obeys the relation (see also \eref{Oz_total})
\begin{eqnarray}
 \label{eq_rot}
 \frac{d \psi}{d t} &= \Omega \quad 
 %\overset{\eref{Oz_total}}{\Rightarrow} 
 \Rightarrow
 \quad 
 \frac{d \psi}{d t} =  A \Omega_0 \sin\psi \nonumber\\
 &{\Rightarrow}~\psi(t) = 2\,\mathrm{arctan}\left[e^{A \Omega_0 t} 
 \tan\left(\frac{\psi(0)}{2} \right)\right]
 \quad
 \end{eqnarray}
with 
\begin{equation}
\label{A_def}
 A: = \frac{1}{8} \left(\frac{R}{h}\right)^3 
 \left[-a_1 + \frac{3}{8} a_3 \left(\frac{R}{h}\right)^2 \right]\,.
\end{equation}
This result has the following implications:\newline
\textbf{(i)} The rotation of the axis depends only on the slope of the gradient 
of the osmotic mobility $b_w$, but not on any particular value of $b_w$ such as 
$b_w(\boldsymbol{\rho}_0$). Therefore the dynamics of the orientation of $\mathbf{d}$ with 
respect to the direction of $\nabla_{||} b_w$ does not depend on the initial 
position $\boldsymbol{\rho}_0$ of the center of the particle.\newline
\textbf{(ii)} The angular velocity $\Omega$ vanishes at $\psi = ~\psi_\infty = 
0,\,\pi$. Therefore the 
rotation of the axis (starting from a typical orientation $\psi(0) \neq 0,\pi$) is 
such that the 
asymptotic ($t \to \infty$) orientation of $\mathbf{d}$ is aligned with the direction of 
$\nabla_{||} b_w$, i.e., thigmotaxis emerges. For $A < 0$, the asymptotic value is 
$\psi_\infty = 0$, 
while for $A > 0$ it is $\psi_\infty = \pi$; therefore $\mathbf{d} \to \mathbf{e}_x$ 
and $\mathbf{d} \to -\mathbf{e}_x$, respectively.\newline
\textbf{(iii)} In \textit{leading} order in $R/h$, the angular velocity $\Omega$ 
carries $a_1$ as a prefactor (see \eref{Oz_total}) which is the contribution 
of the dipole term in the expansion of the density (\eref{c_as_series}). 
For an active 
colloid with a phoretic mobility $b(\mathbf{r}_P)$ which is approximately 
constant over the whole surface, this coefficient $a_1$ determines its 
velocity and thus its motile character in an unbounded fluid (see 
the corresponding expression for $\mathbf{V}^{(b)}$ above \eref{def_V0}). 
Therefore, if an active particle with such $b$ exhibits motility when suspended 
in an unbounded fluid, i.e., $|\mathbf{V}^{(b)}| \neq 0$ and thus $a_1 \neq 0$, 
it will also exhibit thigmotaxis, i.e., $\Omega \neq 0$ for $\psi \neq 0, \pi$, 
if it is near a responsive wall with chemical gradients. In this sense, the 
emergence of thigmotaxis is a generic feature of chemically active colloids which 
are motile in an unbounded fluid. In other words, it is sufficient to know that 
in an unbounded fluid the colloid is motile -- without knowing any other 
details about the mechanism (such as activity function) underlying the motion -- 
in order to be able to predict that the same colloid will exhibit thigmotaxis when 
it is near a responsive wall with surface-chemistry gradients.\newline
\textbf{(iv)} Under the same assumptions as in \textbf{(iii)}, to leading 
order in $h/R$ the ratio of the angular velocity $\Omega$ and the translational 
velocity $V^{(b)}$ in an unbounded fluid is given by (for $a_1 \neq 0$)
\begin{equation}
 \label{ratio_vels}
 \frac{\Omega}{V^{(b)}} = -\frac{3}{16} \frac{\Delta}{b} \left(\frac{R}{h}\right)^3\,.
\end{equation}
This ratio depends only on unexpectedly few material parameters and on the height 
to radius ratio. Most importantly, it is independent of any details of the 
activity function.

\subsection{\label{model_activities} Dependence of the alignment with the 
wall pattern on the details of the chemical activity of the particle}

Here, the general results obtained in the previous section will be discussed for 
three choices of the activity function $f(\theta)$ (see \eref{def_act}). These  
choices are typically employed in studies of chemically active colloids 
\cite{Golestanian2007}:
\numparts
\begin{equation}
\label{act_func_pi}
f^{(up)} = \cases{ 0, & $0 \leq \theta < \frac{\pi}{2}$\\
+1, & $\frac{\pi}{2} \leq \theta \leq \pi$\\}\,,
\end{equation}
\begin{equation}
\label{act_func_vt}
f^{(vp)} = \cases{0 & $0 \leq \theta < \frac{\pi}{2}$\\
- \cos \theta' & $\frac{\pi}{2} \leq \theta \leq \pi$\\}\,,\quad
\end{equation}
\begin{equation}
\label{act_func_pa}
f^{(pa)} = \cases{-1 & $0 \leq \theta < \frac{\pi}{2}$\\
+1 & $\frac{\pi}{2} \leq \theta \leq \pi$\\}\,.\quad
\end{equation}
\endnumparts
The first two choices with superscript $(up)$ and $(vp)$, respectively, 
describe a particle with \textit{u}niform \textit{p}roduction ($(up)$) or 
with a \textit{v}ariable \textit{p}roduction ($(vp)$) of a solute over the 
lower hemisphere (see \fref{fig1}(a)) while the upper hemisphere is chemically 
inactive (thus the particle is indeed a net producer of solute). 
The third choice with superscript $(pa)$ describes a particle which 
\textit{p}roduces a solute with a uniform rate $Q$ over the lower hemisphere 
and \textit{a}nnihilates it, at the same rate $Q$, over the upper 
hemisphere. Thus the particle is not a net producer of solute 
\footnote{All three choices are simple models of experimental realizations of active colloids. 
The first two correspond to the case of, e.g., a platinum (Pt)/silica 
Janus colloid decomposing peroxide (H$_2$O$_2$), with presumably only the 
Pt side being involved in the reaction. The third choice corresponds to 
the situation of a bi-metallic Janus colloid, e.g., Pt/gold(Au), decomposing 
H$_2$O$_2$ via 
oxidation/reduction mechanisms, for which both parts of the colloid 
surface contribute. (Pt is the production side, Au the annihilation side.) 
The reaction 
rate may be taken as spatially uniform (first and third choice) or as dependent 
on the spatial position due to, e.g., a dependence on the thickness 
of the catalyst layer \cite{Golestanian2012,brown14} (second choice). In all three 
cases the Damk\"ohler number is considered to be small, i.e., 
the diffusion-reaction behavior of the H$_2$O$_2$ belongs to the reaction-limited 
regime.}. 
We note that in order for the model corresponding to the third choice to be 
well-defined, i.e., the density (\eref{c_as_series}) to be positive everywhere, 
the bulk density 
$c_\infty$ should be sufficiently large. For these activity 
functions, the first four coefficients $a_n, n = 0,\dots,3$ take the following 
values:
\numparts
 \begin{equation}
  \label{mon_dip_coeff_net}
  a_0^{(up)} = \frac{1}{2}\,,~~a_1^{(up)} = -\frac{3}{8}\,,~~a_2^{(up)} = 
0\,,~~a_3^{(up)} = \frac{7}{64}\,;
 \end{equation}
 %%%%%%
 \begin{equation}
  \label{mon_dip_coeff_var}
  a_0^{(vp)} = \frac{1}{4}\,,~~a_1^{(vp)} = -\frac{1}{4}\,,~~a_2^{(vp)} = 
  \frac{5}{48}\,,~~a_3^{(vp)} = 0\,;
 \end{equation} 
and
\begin{equation}
\label{mon_dip_coeff_nonet}
  a_0^{(pa)} = 0\,,~~a_1^{(pa)} = -\frac{3}{4}\,,~~a_2^{(pa)} = 0\,,~~a_3^{(pa)} = 
\frac{7}{32}\,.
\end{equation}
\endnumparts
We remark that for the first and the third choice all coefficients with an 
even index $n > 0$ are zero, while for the second choice all coefficients 
with an odd index $n > 1$ are zero. The third choice has no monopole 
contribution, which is present in the first two choices. The second choice 
includes a quadrupole contribution ($a_2^{(vp)} \neq 0$), which is missing in the 
other two cases. 

For these three cases, up to the first two orders in $R/h$ the factor $A$, 
\eref{A_def}, is given by
\begin{equation*}
 A^{(up)} = \frac{3}{64} \left(\frac{R}{h}\right)^3 
 \left[1- \frac{7}{64} \left(\frac{R}{h}\right)^2 \right] > 0\,, 
 \qquad A^{(pa)} = 2 A^{(up)} > 0\,,
\end{equation*}
and (\eref{A_def} and \eref{mon_dip_coeff_var})
\begin{equation}
\label{A_results}
 A^{(vp)} = \frac{1}{32} \left(\frac{R}{h}\right)^3  > 0\,.
\end{equation}
Therefore, in each of the three cases the response from the wall aligns the 
colloid such that $\mathbf{d} = - \mathbf{e}_x$, i.e., against the direction 
of the gradient. This corresponds to a configuration of the colloid 
such that its solute producing side faces larger values of $b_w$. This 
behavior is expected intuitively and is consistent with the findings of the 
study in \cite{Uspal2016}. From this analysis we conclude that the thigmotaxis 
phenomenology depends only weakly (via quantitative differences between the 
corresponding angular velocities) on the precise details of the chemical 
activity and thus the motility mechanism.

The relevance of the rotation towards the surface-chemistry gradient, 
characterized by $\Omega$, can be estimated by comparing it with the rotational 
diffusion of the colloid, characterized by the rotational diffusion 
coefficient $D_R$. To this end, for reasons of simplicity we discuss only the case 
(\textit{vp}). Furthermore, for the purpose of a qualitative comparison we use the 
expression $D_R = (3/4) D_T R^{-2}$ \cite{Howse2007} (where $D_T$ denotes the 
translational diffusion coefficient of the particle) for the rotational 
diffusion coefficient of the spherical particle in an unbounded fluid. For a 
Janus particle with $R = 1~\mathrm{\mu m}$ and with a uniform phoretic mobility 
$b$, which in an unbounded fluid moves with 
velocity $V^{(b)} \sim 5~\mathrm{\mu m/s}$ (see \sref{free_space} and the discussion 
after \eref{def_V0}), the translational diffusion coefficient of the particle 
in water at room temperature (with 
viscosity $\mu = 10^{-3}~ \mathrm{Pa} \times \mathrm{s}$) is $D_T = 
k_B T/(6 \pi \mu R) \approx 10^{-13} \mathrm{~m^2/s}$ and 
thus the translational P{\'e}clet number of the  particle is 
$\mathrm{Pe_P} = V^{(b)} R/D_T \approx 50$. The rotational P{\'e}clet 
number of the particle, which we define as $\mathrm{Pe_R} := \Omega/D_R$, 
corresponding to the maximum  magnitude of the angular velocity (attained at 
$\psi = \pi/2$, see \eref{Oz_total}) is (\eref{A_results}, \eref{char_Oz}, 
\eref{def_V0}, and $\mathbf{V}^{(b)}$ as given at the end of \sref{free_space})
\begin{eqnarray}
\label{Pe_rot}
 \mathrm{Pe_R} &:= \frac{A^{(vp)} \Omega_0}{D_R} 
 = \frac{1}{32} \left(\frac{R}{h}\right)^3 \frac{Q \Delta}{D}
 \frac{4}{3} \frac{R^2}{D_T} \nonumber\\
 & = \frac{1}{16} \frac{1}{|a_1^{(vp)}|} \left(\frac{R}{h}\right)^3 
 \frac{R \Delta}{|b|} \frac{V^{(b)} R}{D_T} 
 = \frac{1}{4} \left(\frac{R}{h}\right)^3 \frac{R \Delta}{|b|} \mathrm{Pe_P}\,.
\end{eqnarray}
This implies that for, e.g., $h = 2 R$ one has $\mathrm{Pe_R} \approx 
1.6\, (R \Delta/|b|)$. Therefore the rotation of $\mathbf{d}$ towards the 
direction of the surface-chemistry gradient, as induced by the response 
of the wall, dominates 
over the rotational diffusion (caused by the thermal fluctuations of the 
orientation of the particle) in regions with steep variations of the surface 
chemistry, i.e., for $\Delta \gg |b|/R$; e.g., the effect is very pronounced near 
sharp chemical steps as studied in \cite{Uspal2016}. This condition may be 
significantly relaxed if the particle is constrained to move at smaller 
distances $h$ from the wall, which increases $\mathrm{Pe_R}$. However, 
in such cases the validity of the 
point-particle analysis presented here must be carefully cross-checked against 
full numerical solutions. Finally, we note that if the surface-chemistry 
gradients fulfill $\Delta \approx |b|/R)$, rotational diffusion and 
deterministic aligning are competing effects of similar 
magnitude. Therefore, in this case we expect the motion to exhibit noisy oscillations 
around a state which appears to be aligned on average. 

\section{\label{conc}Summary and conclusions}

We have shown that a chemically active colloid, which moves in 
the vicinity of a chemically heterogeneous wall, can detect these variations in the 
chemical composition of the wall via induced osmotic flows which couple 
back to the colloid. For a constant surface-chemistry gradient along the wall, 
the colloid exhibits a tendency to align its axis with the direction of 
this gradient. This phenomenon, which is a primitive form of guidance 
by sensing the environment, resembles that of thigmotaxis, i.e., a response to 
the proximity of a physical change in the environment as exhibited by living 
organisms. 

By employing a point particle approximation (\sref{set-up wall motion} and 
\sref{sec_general_response}) we have characterized this phenomenon by 
calculating the  angular velocity of the rotation of the axis of the particle 
induced by the osmotic flows generated at the wall (\eref{Oz_form2} 
and \eref{wall_dens}). This is determined in leading order 
(\eref{Oz_total}), by the ``dipole'' coefficient $a_1$ 
(\eref{c_as_series} and \eref{multipole_mag}), of the 
chemical activity. This coefficient also determines the motility of the colloid 
in an unbounded 
fluid, under the proviso that the phoretic mobility coefficient $b$ (see 
\eref{def_b}) is uniform over its surface (see \sref{free_space}). This allows 
us to conclude that the emergence of thigmotaxis, i.e.,  $\Omega \neq 0$, is a 
generic feature for active colloids which are motile in an unbounded fluid, 
which requires $a_1 \neq 0$ (see above \eref{def_V0}). 

By using several simple models of active colloids, in \sref{model_activities} 
we have shown (see \eref{A_results}) that the thigmotactic alignment depends 
only weakly on the precise details of the chemical activity and thus of the 
motility mechanism. Finally, we have estimated (\eref{Pe_rot}) the 
magnitude of the surface-chemistry gradients needed so that the tendency for 
thigmotaxis dominates over the thermal fluctuations of the orientation 
$\mathbf{d}$ of the colloid in a plane parallel to the wall. 

Once $\mathbf{d}$ is aligned with the direction $\nabla_{||} b_w$ of the 
surface-chemistry gradient at the wall, the distribution of the solute number 
density at the wall (see \fref{fig2}) and the response of the wall (via the 
chemi-osmotic 
coefficient $b_w$ (\eref{def_osmo_b})) have mirror-symmetry with 
respect to the plane which includes $\mathbf{d}$ and is normal to the wall 
(see \fref{fig2}). This ensures that a wall-induced translation along the 
$y$-direction, i.e., perpendicular to the surface-chemistry gradient 
(see \fref{fig1}(b)), vanishes in this state, i.e., $V^{(w)}_y = 0$. The 
translation $V_x$ along the direction of the surface-chemistry gradient depends on all 
the details of the wall chemistry, including, via $V^{(w)}_x$ (see 
\eref{Vx_form1}), the local value $b_w(\boldsymbol{\rho}_0)$ of the osmotic 
mobility coefficient at the lateral position $\boldsymbol{\rho}_0$ of the colloid, 
as well as, via $V^{(sd)}$ (see \sref{set-up wall motion}, first sub-problem), 
on the details of the motility mechanism of the particle. Therefore this aspect was 
not pursued here further.

Finally we note that the emergence of thigmotaxis, as studied here, raises the 
question of whether this can provide an effective mechanism for ``source detection'' 
problems, e.g., if the surface-chemistry gradient of the wall is not constant 
but rather decays with the distance from a point or a line. This generalizes the 
case of step-like variations in $b_w$ which were studied in Ref. 
\cite{Uspal2016}.

\ack
M.N.P. and W.E.U. acknowledge financial support from the German Science 
Foundation (DFG), grant no. TA 959/1-1.

\appendix

\section*{References}

\bibliography{refs}

\end{document}